\documentclass[aps,pre,groupedaddress,twocolumn,showpacs]{revtex4}
\usepackage{graphicx}
\newcommand\mean[1]{{\langle#1\rangle}}

\newcommand{\Ocal}{{\cal O}}
\newcommand{\Ccal}{{\cal K}}

\begin{document}

\title{Hybrid method for simulating front propagation in
  reaction-diffusion systems}

\author{Esteban Moro}
\email[]{emoro@math.uc3m.es}
\homepage[]{http://gisc.uc3m.es/~moro}
\affiliation{
Grupo Interdisciplinar de Sistemas Complejos (GISC) and
Departmento de Matem\'aticas, Universitas Carlos III de Madrid,
Avda.\ de la Universidad 30, E-28911 Legan\'es, Spain
}%

\date{\today}

\begin{abstract}

We study the propagation of pulled fronts in the $A
\leftrightarrow A+A$ microscopic reaction-diffusion process using
Monte Carlo (MC) simulations. In the mean field approximation the
process is described by the deterministic
Fisher-Kolmogorov-Petrovsky-Piscounov (FKPP) equation. In
particular we concentrate on the corrections to the deterministic
behavior due to the number of particles per site $\Omega$. By
means of a new hybrid simulation scheme, we manage to reach large
macroscopic values of $\Omega$ which allows us to show the
importance in the dynamics of microscopic pulled fronts of the
interplay of microscopic fluctuations and their macroscopic
relaxation.

\end{abstract}


\maketitle

When describing systems at much larger scales than correlation
length, internal fluctuations due to the intrinsic discreteness of
the particles can be neglected, since they only account for a
correction typically of the order of $\Omega^{-1/2}$, where
$\Omega$ is the number of particles in a correlated volume.
However, in some situations the dynamics of the system spans
different scales which give rise to a strong dependence of its
macroscopic features on the microscopic details of its
constituents, even in the limit $\Omega \to \infty$. Relevant
instances of this phenomena are the dynamic contact angle problem
\cite{angle}, evolution of a fracture tip \cite{fracture},
dendritic growth \cite{karma}, and the flow of a gas through a
microscopic channel \cite{garcia}. In this paper we highlight and
study another important example namely, the effect of internal
fluctuations in the macroscopic dynamics of pulled fronts
\cite{vansaarloos,panja}. Specifically we consider the propagation
of pulled fronts in reaction-diffusion microscopic problems, like
the $A \leftrightarrow A + A$ scheme \cite{breuer1,avraham,moro1}.
Continuum description of the system is possible in the
reaction-limited regime where $\Omega$ is large enough so the
reaction is well stirred within each correlated volume
\cite{avraham,moro1}. In this case, the density $\rho(x,t)$ of
particles per correlated volume is described, in the limit
$\Omega\to\infty$, by the FKPP equation \cite{fisher}
\begin{equation}\label{fkpp}
\frac{\partial \rho}{\partial t} = D \frac{\partial^2
\rho}{\partial x^2} + k_1 \rho - k_2 \rho^2.
\end{equation}
This equation has travelling-wave solutions of the form $\rho =
\rho(x-vt)$ which invade the unstable phase $\rho(\infty)=0$ from
the stable phase $\rho(-\infty) = k_1/k_2$ and travel with
velocity $v\geq v_0 = 2 \sqrt{D k_1}$. Of particular interest is
the solution with velocity $v_0$, since it is dynamically selected
for a broad class of initial conditions. Moreover, $v_0$ is the
linear spreading speed of infinitesimal perturbations around the
unstable state. Thus, fronts with velocity $v_0$ are essentially
``pulled along'' by the growth and spreading of small
perturbations in the leading edge $x \gg vt$ where $\rho \ll 1$.
This sensitivity causes also the {\em absence of a typical
macroscopic length and time scale} in which perturbations around
the asymptotic solution with velocity $v_0$ are damped
\cite{vansaarloos}. For example, the velocity of the front
starting from an steep enough initial condition approaches the
asymptotic value like a power law
\begin{equation}\label{velocity}
v(t) = v_0 - \frac{3}{2 q_0 t} + \Ocal(t^{-3/2}),
\end{equation}
where $q_0 = v_0/2D$.

In particle models however, the continuum description given by the
FKPP equation breaks down at $\rho \simeq 1/\Omega$ where internal
fluctuations are important. Since pulled front dynamics are
sensitive to infinitesimal events at $\rho \ll 1$ we expect
macroscopic properties to depend strongly on $\Omega$ when $\Omega
\to \infty$. For example, by neglecting microscopic fluctuations
and mimicking the discreteness of particles by imposing an
effective cutoff in the FKPP equation at $\rho = \Omega^{-1}$,
Brunet and Derrida \cite{brunet1,brunet2} obtained that the
velocity is given by
\begin{equation}\label{bd1}
v(\infty) \equiv v_\Omega = v_0 - \frac{v_0 \Ccal_v}{\ln^2 \Omega}
+ \Ocal(\ln^{-5/2} \Omega).
\end{equation}
where $\Ccal_v$ is a constant. As expected the correction to the
macroscopic velocity of the front is very strong: in a macroscopic
volume of $10^{23}$ particles, it is still of $0.3 \%$. Combining
Eqs.\ (\ref{velocity}) and (\ref{bd1}) we can easily infer that
the typical time scale of microscopic pulled fronts is given by
the condition $v(\tau_\Omega) \simeq v_\Omega$, that is
\cite{kessler}
\begin{equation}
\tau_\Omega \sim \ln^2 \Omega
\end{equation}
Thus, pulled fronts in reaction diffusion particle models do have
a typical time scale as opposite to those of the FKPP equation,
although it is set by microscopic details and diverges in the
limit $\Omega \to \infty$ \cite{footnote2}.

Numerical confirmation of these predictions in general
reaction-diffusion particle models is difficult, since the
observation of the functional dependence in (\ref{bd1}) requires
typical simulations up to $\ln \Omega \simeq 10^2$ which are
computationally unfeasible. However, for a particular model in
which particles undergo non-local diffusion movements, Brunet and
Derrida where able to perform simulations up to $\Omega \simeq
10^{150}$ and check with high accuracy the prediction (\ref{bd1})
\cite{brunet1,brunet2}. Moreover, they found also that the front
diffuses in time and that the diffusion coefficient behaves as
\cite{brunet2}
\begin{equation}\label{bd2}
D_\Omega \simeq \frac{\Ccal_D}{\ln^3 \Omega}.
\end{equation}
where $\Ccal_D$ is a constant. While the velocity correction
(\ref{bd1}) can be easily understood in terms of an effective
cutoff in the FKPP equation \cite{brunet1}, and simulations for
moderate numbers of particles ($\Omega \leq 10^{10}$) in
reaction-diffusion particle models {\em seem to be compatible}
with (\ref{bd1}) \cite{kessler,lemarchand,panja,moro1}, the
functional dependence of $D_\Omega$ has only been observed in the
non-local model of Brunet and Derrida. Although there is an
heuristic argument for a specific model to get the $\ln^{-3}
\Omega$ dependence \cite{panja}, the situation clearly remains
unsatisfactory, since there is only one empirical observation of
(\ref{bd2}). Thus, our purpose in this paper is to simulate the
$A\leftrightarrow A+A$ model for very large number of particles in
order to check both (\ref{bd1}) and (\ref{bd2}) and get some
insight into the dynamics of pulled fronts in microscopic particle
models.

The $A\leftrightarrow A+A$ model in one dimension consists of
particles on a lattice with spacing $\Delta x$ in which the number
of particles at site $i$, $N_i(t)$ is unbounded \cite{breuer1}.
Reaction events take place on-site, while diffusion drives
particles to nearest neighbors positions. Particles annihilate
with rate $\sigma$, give birth to another one with rate $\gamma$
and diffuse with rate $D$. In equilibrium, the average number of
particles per site is $\Omega = \gamma / \sigma$, and when $\Omega
\to \infty$ the system is described by Eq.\ (\ref{fkpp}) with
$\rho(x,t) = N(x,t)/\Omega$ and $k_1=k_2=\gamma$. Early MC
simulations of this model \cite{breuer1} showed that indeed when
$\Omega \gg 1$, FKPP pulled fronts emerge. Specifically, it was
found that both the velocity correction and the diffusion
coefficient of the front decay like $\Omega^{-1/3}$ for $\Omega
\leq 10^6$, a scaling which has been observed in other models for
moderate values of $\Omega$ \cite{lemarchand}. However, in order
to observe the scalings (\ref{bd1}) and (\ref{bd2}) much larger
numbers of particles are needed (typically $\Omega \gg 10^{10}$)
which can not be attained in standard MC simulations.

\begin{figure}
\begin{center}
\includegraphics[width=3.2in,clip=]{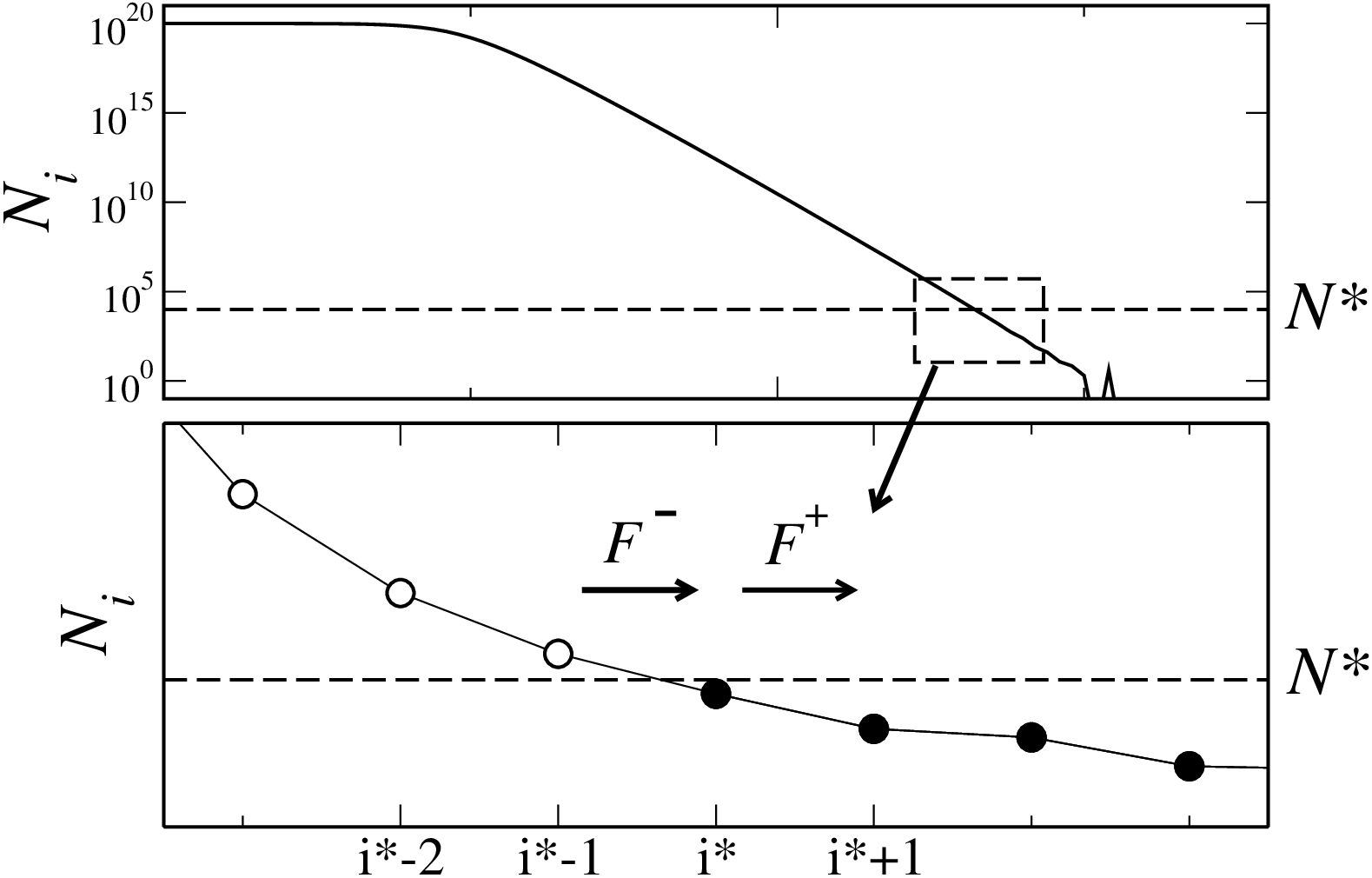}
\caption{\label{fig1} Snapshot of the front profile for one
realization of the hybrid method with $\Omega = 10^{20}$ (upper
panel). Below: schematic view of the front $N_i(t)$ close to
$i^*$. Open symbols are in the macroscopic region and full symbols
in the microscopic region. For each point lattice $i$ we define
the flux of outgoing $F^+_i(t)$ and incoming $F^-_i(t)$ particles.
In the figure only those for $i^*$ are shown. Lines are guides to
the eye. }
\end{center}
\end{figure}

To reach larger number of particles in the $A \leftrightarrow A+A$
model we note that, since the dynamics of pulled fronts are very
sensitive to the dynamics of the system close to the unstable
state $\rho \simeq 0$, a correct description of the
reaction-diffusion microscopic problem is only needed there, where
fluctuations are important \cite{brunet2}. Away from the unstable
state, i.e.\ when $\rho = \Ocal(1)$, the number of particles is
big enough so that fluctuations are negligible and the system can
be safely described by macroscopic descriptions like (\ref{fkpp}).
Thus we propose to split the dynamics of the microscopic model
into two different descriptions: given a mesoscopic number of
particles $N^*$ with $\Omega \gg N^* \gg 1$, at any time step $t$
we identify the position $i^*$ as the smallest value of $i$ for
which $N_i(t) \leq N^*$. In the region in which fluctuations can
be safely ignored, that is when $i < i^*$ we update the number of
particles using a numerical approximation of Eq.\ (\ref{fkpp}),
while we use MC methods in the region $i \geq i^*$. To complete
the algorithm, boundary conditions at $i = i^*$ should be given.
Since only diffusion couples the dynamics between different sites,
we implement the boundary condition through the conservation of
fluxes of particles through the boundary, similarly to other MC
hybrid methods \cite{garcia1,flekkoy,smekera}.

To this end, if we define at each site $i$ the flux of incoming
particles $F^-_i(t)$ and of outgoing particles $F^+_i(t)$ (see
Fig.\ \ref{fig1}), the Euler approximation with time step $\Delta
t$ of Eq.\ (\ref{fkpp}) reads
\begin{equation}
\frac{N_i(t+\Delta t)-N_i(t)}{\Delta
t}=\frac{F^-_i(t)-F^+_i(t)}{\Delta x} +  \gamma N_i(t) - \sigma
N_i^2(t). \label{flux}
\end{equation}
with
\begin{eqnarray}
F_i^-(t)&=&\frac{D}{\Delta x}[N_{i-1}(t)-N_i(t)] \nonumber \\
F_i^+(t)&=&\frac{D}{\Delta x} [N_i(t)-N_{i+1}(t)]\label{fluxes}
\end{eqnarray}
Obviously, conservation of the number of particles requires that
$F^-_i(t) = F^+_{i-1}(t)$ and $F^+_i(t)=F^-_{i+1}(t)$. Thus, our
algorithm evolves as follows: for a given mesoscopic time step
$\Delta \tilde t$, we update the microscopic region ($i \geq i^*$)
using time continuous MC \cite{breuer1} until the time of the
simulation $\Delta t$ exceeds $\Delta \tilde t$. Note that the
typical time step in the MC simulation is given by $\delta t^{-1}
\simeq N^*[1+\log(\Omega/N^*)]$ \cite{footnote1}, which is smaller
than $\Delta \tilde t$ in our simulations. Thus, several MC events
take place until the MC simulation time $\Delta t$ exceeds $\Delta
\tilde t$, which makes the real $\Delta t$ different for any time
step. Any MC event in which a particle jumps into the macroscopic
region ($i < i^*$) is recorded in the variable $N^-$ and the
particle is removed from the MC simulation. We then update sites
$i \leq i^*-2$ using Eq.\ (\ref{flux}) and (\ref{fluxes}).
Finally, the number of particles at the boundary site $i^*-1$ is
updated using equation (\ref{flux}) but with $F^+_{i^*-1}$
calculated according to the MC recorded number of jumps, $N^-$.
Specifically we take $F^+_{i^*-1} = (D/\Delta x) N_{i^*-1}(t) -
N^-/(\Delta t \Delta x)$. Since we should get that $F^-_{i^*}(t) =
F^+_{i^*-1}(t)$, we update the number of particles at site $i^*$
to fulfill this condition on average: $N_{i^*}(t+\Delta t) =
N_{i^*}(t+\Delta t) + \Pi_{\Delta t D N_{i^*-1}(t)/\Delta x}$
where $\Pi_\lambda$ is a Poisson random number with mean
$\lambda$. This completes a time step $\Delta t$ in the algorithm.

The condition for a sharp interface between the macroscopic and
the microscopic region at $i^*$ can be relaxed by introducing a
buffer region \cite{garcia1,flekkoy}. Moreover, fluctuations can
be also considered in the macroscopic region by adding an internal
noise source to the FKPP equation \cite{sfkpp,garcia1}. However,
for large enough $N^*$ and small enough $\Delta \tilde t$ our
results do not differ from those of these algorithm refinements.
In our simulations we take $\Delta \tilde t = 10^{-4}$, $N^* =
\min\{10^4,\Omega/2\}$ and $D = \gamma = \Delta x = 1$.

\begin{figure}
\begin{center}
\includegraphics[width=3.2in]{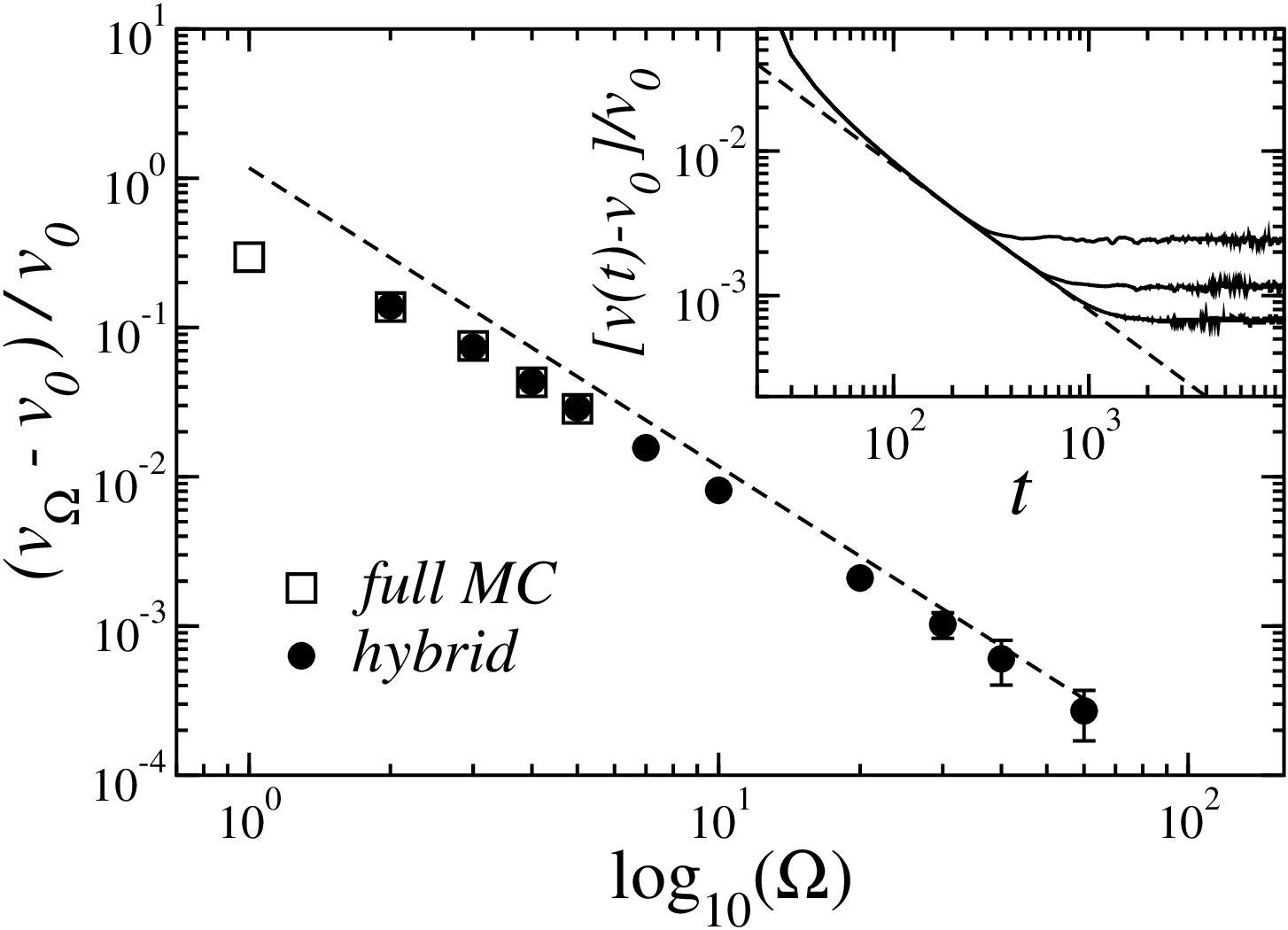}
\caption{\label{vel} Asymptotic velocity correction as a function
of the logarithm of the number of particles $\Omega$. Open symbols
are results for the full MC simulation while full symbols are for
the hybrid scheme. Dashed line corresponds to the scaling given by
(\ref{bd1}) with $\Ccal_v$ given by (\ref{Ccal}). Error bars are
not shown when smaller than the symbol size. Inset: time evolution
of the instantaneous velocity of the front (solid lines) with
$\Omega = 10^{20}, 10^{30}$ and $10^{40}$ from top to bottom.
Dashed line is the prediction given by Eq.\ (\ref{velocity}).}
\end{center}
\end{figure}

\begin{figure}
\begin{center}
\includegraphics[width=3.2in,clip=]{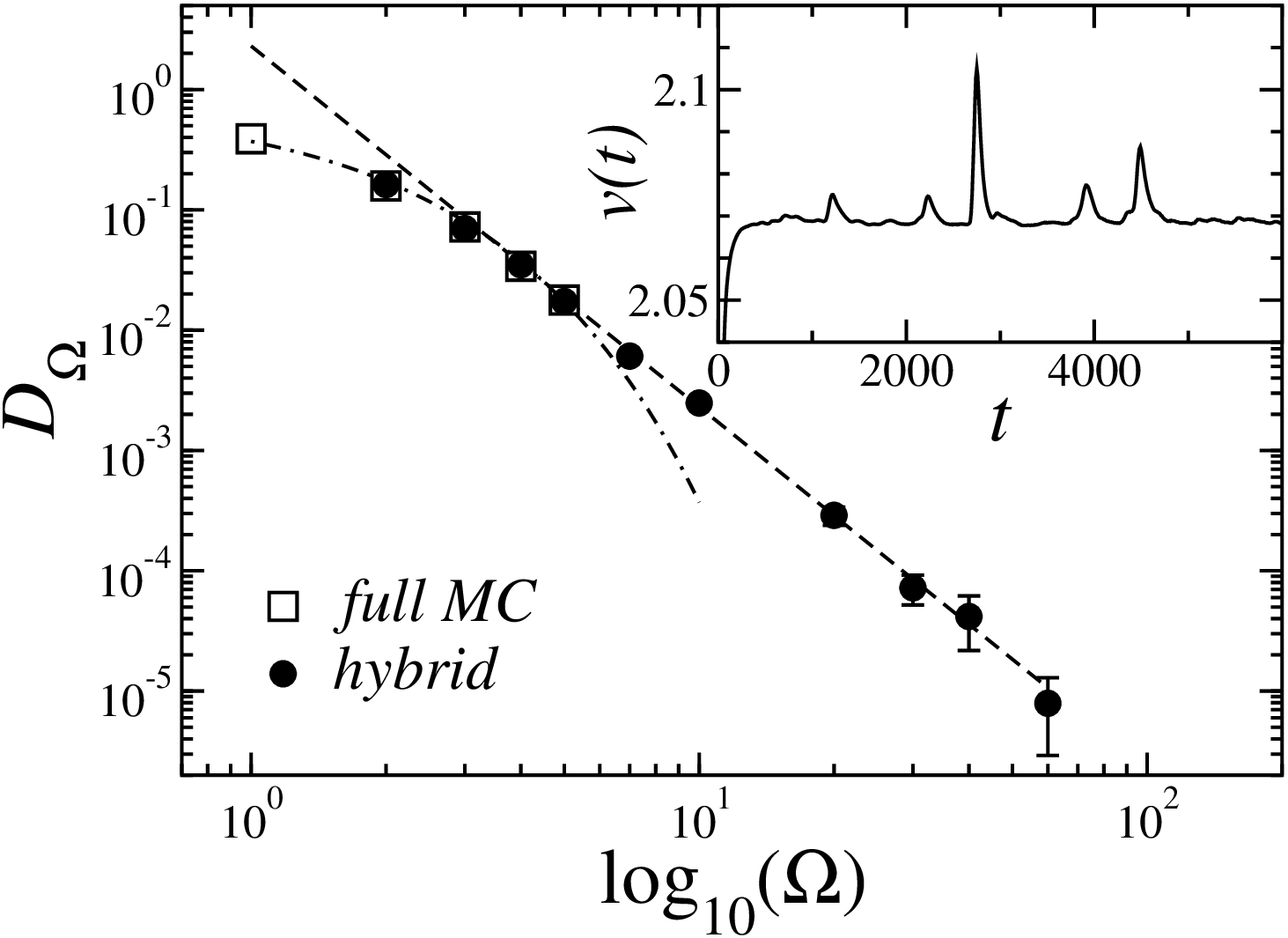}
\caption{\label{dif} Diffusion coefficient of the front $D_f$ as a
function of the logarithm of the number of particles $\Omega$.
Symbols are as in figure \ref{vel}. Dashed line is a fit of the
last points to the scaling form (\ref{bd2}) with $\Ccal_D = 26.5$.
Dotted-dashed line is the scaling $D_f \sim \sigma^{-0.32}$ of
\cite{breuer1}. Error bars are not shown when smaller than the
symbol size. Inset: Velocity as a function of time for one
realization of the algorithm with $\Omega = 10^{20}$.}
\end{center}
\end{figure}

Results for this hybrid scheme are shown in figures \ref{vel} and
\ref{dif} and compared with full MC simulations up to $\Omega =
10^5$. Both the correction to the velocity of the front
(\ref{bd1}) and the diffusion coefficient (\ref{bd2}) agree, up to
statistical fluctuations, with those of the full MC simulations,
which supports the validity of our algorithm. Note that when
$\Omega \to \infty$ our algorithm reduces to the Euler
approximation of the FKPP equation. Thus, the velocity $v_0$ in
(\ref{bd1}) is given by the solutions of the equations
\cite{pechenik}
\begin{eqnarray}
v_0 e^{-q_0 v_0 \Delta \tilde t}&=&-2 \sinh q_0, \nonumber \\
e^{-q_0 v_0 \Delta \tilde t}&=&2 \Delta \tilde t\: [ (\cosh q_0
-1) -1], \label{eqpechenik}
\end{eqnarray}
and the velocity correction coefficient obtained using the
efficient deterministic cutoff argument of \cite{brunet1} is given
by
\begin{equation}\label{Ccal}
\Ccal_v = \pi^2 q_0 (e^{v_0 q_0 \Delta t}\cosh q_0 - v_0^2 \Delta
t/2).
\end{equation}

In the limit $\Omega \to \infty$ we observe in Fig.\ \ref{vel}
that our results tend to the scaling (\ref{bd1}) together with the
solutions (\ref{eqpechenik}) and (\ref{Ccal}). However, a strong
deviation of our results for the predicted scaling (\ref{bd1})
even for large values of $\Omega$ is observed, a fact that also
was present in the Brunet and Derrida model \cite{brunet2}.
Asymptotic convergence of microscopic pulled fronts in the $A
\leftrightarrow A+A$ towards the solution of the FKPP equation is
also observed in the inset of Fig.\ \ref{vel} in which we plot the
time dependence of the velocity for different values of $\Omega$.
As expected, in our simulations the velocity decays accurately
like (\ref{velocity}) until it saturates to a constant value given
by (\ref{bd1}).

\begin{figure}
\begin{center}
\includegraphics[width=3.2in,clip=]{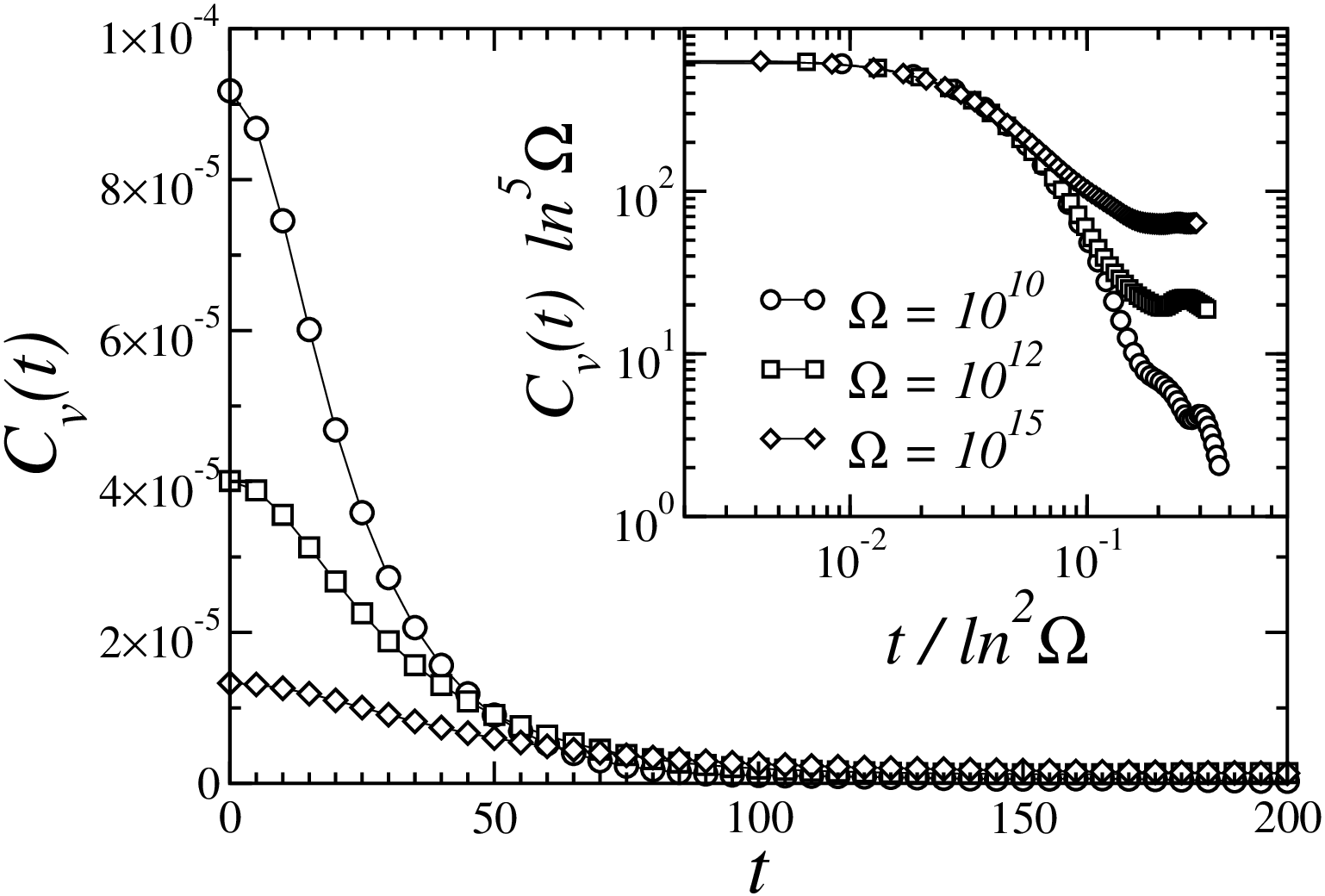}
\caption{\label{corr} Velocity time correlation as a function of
time for different values of the number of particles $\Omega$.
Inset shows the scaling given by Eq.\ (\ref{scaling}).}
\end{center}
\end{figure}

Regarding the diffusion coefficient, our results for the $A
\leftrightarrow A + A$ confirm the scaling (\ref{bd2}) found in
\cite{brunet2}. Note that the results for small values of $\Omega$
agree with the scaling $D_\Omega \sim \Omega^{-1/3}$ found in the
initial studies of the $A \leftrightarrow A+A$ model
\cite{breuer1}. To understand the origin of the functional
dependence of the diffusion coefficient, we show in the inset of
figure \ref{dif} a typical realization of the instantaneous
velocity of the front as a function of time. As we can see,
long-lived fluctuations take place at the front, whose origin is
in the large relaxation time of microscopic pulled front dynamics,
$\tau_\Omega \sim \ln^2\Omega$. In order to check this, we have
measured the time correlation of the instantaneous velocity of the
front for different values of $\Omega$:
\begin{equation}\label{velcorr}
C_v(t)\equiv \mean{[v(s+t)-v_\Omega][v(s)-v_\Omega]}.
\end{equation}
Our data (see figure \ref{corr}) indicate that the velocity
correlation scales like
\begin{equation}\label{scaling}
C_v(t) \sim \frac{1}{\ln^{5}\Omega}\: G\left( \frac{t}{\ln^{2}
\Omega}\right),
\end{equation}
where $G(x)$ is a scaling function. As expected, the relaxation
time of velocity fluctuations is given by $\tau_\Omega$ and, in
particular, Eq.\ (\ref{scaling}) is consistent with the scaling of
$D_\Omega$ given by (\ref{bd2}) using the Kubo formula
\begin{equation}\label{Kubo}
D_\Omega \sim \lim_{t,s\to\infty}\int_s^{s+t} C_v(t') dt' \sim
\frac{1}{\ln^3\Omega}.
\end{equation}

The observed scaling (\ref{scaling}) gives us some insight about
the effect of internal fluctuations in microscopic pulled fronts:
at small densities, fluctuations in the number of particles
$N_i(t)$ become important (for example, note in Fig.\ \ref{fig1}
the presence of particles well ahead of the tip of the front).
Equation (\ref{scaling}) suggests that those fluctuations have a
strength proportional to $1/\ln^5 \Omega$. In the existence of a
typical macroscopic scale, those fluctuations would be damped
almost instantaneously and the diffusion coefficient would have
been proportional to $\ln^5 \Omega$. In fact, a similar result is
obtained analytically for the coarse-grained continuous model (the
stochastic FKPP equation \cite{sfkpp}) of the $A\leftrightarrow A
+ A $ model when standard perturbation techniques (which rely in
the existence of a macroscopic time scale) are used \cite{panja}.
However, microscopic pulled fronts do not have this macroscopic
time scale and fluctuations are accommodated by the dynamics in a
much larger time scale, $\tau_\Omega$. The interplay between the
microscopic fluctuations of strength $\ln^{-5} \Omega$ and the
time scale of order $\ln^2 \Omega$ in which they relax is what
produces the dependence with $\ln \Omega$ observed in the
diffusion coefficient.

In summary, we have presented a new hybrid method for studying the
dynamics of fronts in particle reaction-diffusion problems. This
hybrid scheme allow us to investigate the asymptotic convergence
of those microscopic models to the macroscopic description given
by the FKPP equation (\ref{fkpp}). In particular, we reproduced
the scaling of the velocity correction with the number of
particles given by (\ref{bd1}) observed in \cite{brunet1} and
inferred in other works. More interestingly, we have confirmed the
proposed scaling for the diffusion coefficient (\ref{bd2}) and
showed that its origin is in the interplay of the typical
relaxation time of microscopic pulled fronts and the strength of
the microscopic fluctuations at small densities.

We would like to thank R.\ Cuerno, C.\ R.\ Doering, A. S\'anchez
and P.\ Smereka for comments and discussions and the MCTP at
University of Michigan and the DEAS at Harvard University for
their hospitality during the progress of this work. This work has
been supported by the Ministerio de Ciencia y Tecnolog\'{\i}a and
Comunidad de Madrid (Spain) grants.


\begin{thebibliography}{99}
\bibitem{angle} J.\ Koplik, and J.\ R.\ Banavar, Ann.\ Rev.\ Fluid
Mech.\ {\bf 27}, 257 (1995).
\bibitem{fracture} F.\ F.\ Abraham, J.\ Q.\ Broughton, N.\
Bernstein, and E.\ Kaxiras, Comput.\ Phys. {\bf 12}, 538 (1998).
\bibitem{karma} M.\ Plapp and A.\ Karma, Phys.\ Rev.\ Lett.\ {\bf
84}, 1740 (2000).
\bibitem{garcia} F.\ Alexander, A.\ L.\ Garcia, and B.\ J.\ Alder,
Phys.\ Fluids {\bf 6}, 3854 (1994); A.\ L.\ Garcia, J.\ B. Bell,
W.\ Y.\ Crutchfield, and B.\ J.\ Alder, J.\ Comp.\ Phys.\ {\bf
154}, 134 (1999).
\bibitem{vansaarloos} W. van Saarloos, Phys.\ Rep.\ {\bf 386} 29 (2003); U.\ Ebert
and W.\ van Saarloos, Physica D {\bf 146}, 1 (2000).
\bibitem{panja} D.\ Panja,  cond-mat/0307363; Phys.\ Rev.\ E {\bf 68}, 065202
(2003).
\bibitem{avraham} D.\ ben-Avraham and S.\ Havlin, {\em Diffusion and
    Reactions in Fractals and Disordered Systems} (Cambridge
  University Press, Cambridge, 2000).
  \bibitem{breuer1} H.\ P.\ Breuer, W.\ Huber, and F.\ Petruccione,
  Physica D {\bf 73}, 259 (1993); Europhys. Lett. {\bf 30}, 69 (1995).
\bibitem{moro1} E. Moro, Phys.\ Rev.\ Lett.\ {\bf 87}, 238303 (2001); Phys.\ Rev.\ E, {\bf 68}, 025102 (2003).
\bibitem{fisher} R.\ A.\ Fisher, Ann.\ Eugenics {\bf VII}, 355
(1936); A.\ Kolmogorov, I.\ Petrovsky, and N.\ Piscounov, Mosc.\
Univ.\ Bull.\ Math.\ A {\bf 1}, 1 (1937).
\bibitem{brunet1} E.\ Brunet and B.\ Derrida, Phys.\ Rev. E {\bf 56}
  2597 (1997); Comput.\ Phys.\ Commun.\ {\bf 122}, 376 (1999).
\bibitem{brunet2} E.\ Brunet and B.\ Derrida, J. Stat. Phys. {\bf
    103}, 269 (2001).
\bibitem{kessler} D.\ A.\ Kessler, Z.\ Ner, and L.\ M.\ Sander,
Phys.\ Rev.\ E {\bf 58}, 107 (1998).
\bibitem{footnote2} Strictly speaking, only fronts with $\tau_\Omega \to \infty$
are pulled, while for finite and large $\tau_\Omega$ they become
weakly pushed, see \cite{vansaarloos}.
\bibitem{lemarchand} A.\ Lemarchand and B.\ Nowakowski, J.\ Chem.\ Phys.\ {\bf 111}, 6190 (1999);
A.\ Lemarchand, J.\ Stat.\ Phys. {\bf 101}, 579 (2000).
\bibitem{garcia1} F.\ J.\ Alexander, A.\ L.\ Garc\'{\i}a and D.\ M.\
  Tartakovsky, J.\ Comp.\ Phys.\ {\bf 182}, 47 (2002).
\bibitem{flekkoy} E.\ G.\ Flekk\o y, G.\ Wagner and J.\ Feder,
  Europhys.\ Lett.\ {\bf 52}, 271 (2000); Phys.\ Rev.\ E {\bf 64},
  066302 (2001).
\bibitem{smekera} T.\ P.\ Schulze, P.\ Smereka and W. E, J.\
Comp.\ Phys.\ {\bf 189}, 197 (2003).
\bibitem{footnote1} The typical time step in a MC simulation is
  $\delta t \simeq 1/M$, where $M$ is the total number of particles.
  Assuming that the front has the shape $\rho(\xi,t) = \xi e^{-\xi}$
  we can approximate $\delta t^{-1} \simeq M = \Omega
  \int_{x^*}^\infty \rho(\xi,t) d\xi = N^* [1+\log(\Omega/N^*)]$.
  \bibitem{sfkpp} C.\ R.\ Doering, C.\ Mueller, and P.\ Smereka, Physica
  A {\bf 325}, 243 (2003).
\bibitem{pechenik} L.\ Pechenik and H.\ Levine, Phys.\ Rev.\ E
{\bf 59}, 3893 (1999).
\end{thebibliography}
\end{document}